\documentclass[prl,reprint,superscriptaddress, longbibliography]{revtex4-1}
\usepackage{amsmath}
\usepackage[pdftex,colorlinks=true]{hyperref}
\usepackage{amssymb}
\usepackage{amsthm}
\usepackage{amsfonts}
\usepackage{amsthm}
\usepackage{bbold}
\usepackage{graphicx}
\usepackage{mathtools}
\usepackage{enumitem}
\usepackage{bbm}
\usepackage{array}
\hypersetup{
citecolor = blue
}
\newcommand{\floor}[1]{\lfloor #1\rfloor}
\usepackage[normalem]{ulem}

\newcommand{\ssection}[1]{\noindent{{{\bf #1}}}}
\newcommand{\sssection}[1]{\noindent{{ \textbf{\textit{#1}}---}}}

\begin{document}
\title{A comment on ``Discrete time crystals: rigidity, criticality, and realizations''}

\author{Vedika~Khemani}
\affiliation{Department of Physics, Stanford University, Stanford, CA 94305, USA}

\author{Roderich Moessner}
\affiliation{Max-Planck-Institut f\"{u}r Physik komplexer Systeme, 01187 Dresden, Germany}

\author{S.~L.~Sondhi}
\affiliation{Department of Physics, Princeton University, Princeton, New Jersey 08544, USA}

\begin{abstract}
The Letter by N. Y. Yao et. al.~\cite{YaoDTC, YaoErratum} presents three models for realizing a many-body localized discrete time-crystal (MBL DTC): a short-ranged model~\cite{YaoDTC}, its revised version~\cite{YaoErratum}, as well as a long-range model of a trapped ion experiment~\cite{YaoDTC, MonroeExp}. We show that none of these realize an MBL DTC for the parameter ranges quoted in Refs.~\cite{YaoDTC, YaoErratum}.  The central phase diagrams in \cite{YaoDTC} therefore cannot be reproduced. The models show rapid decay of oscillations from generic initial states, in sharp contrast to the robust period doubling dynamics characteristic of an MBL DTC. Long-lived oscillations from special initial states (such as polarized states) can be understood from the familiar low-temperature physics of a {\it static} transverse field Ising model, rather than the  nonequilibrium physics of an eigenstate-ordered MBL DTC. Our results on the long-range model also demonstrate, by extension, the absence of an MBL DTC in the trapped ion experiment of Ref.~\cite{MonroeExp}. 
\end{abstract}
\maketitle

An MBL DTC spontaneously breaks time-translation symmetry, and is a paradigmatic example of a novel non-equilibrium phase in a periodically driven many-body quantum system~\cite{Khemani2016, Else2016, vonKeyserlingk2016}. There is much current interest in realizing a DTC in various laboratory systems~\cite{GoogleTCExp, DelftTCExp, RachelTCExp,  BeijingTCExp}. This has been motivated, in part, by the recognition that various early experiments~\cite{MonroeExp, LukinExp, Sreejith, BarrettExp}, while instructive, did not in fact realize an MBL DTC~\cite{Khemani2019, ippoliti2021manybody}. This renewed effort requires an understanding of the prerequisites for observing the MBL DTC phase, i.e.\ a reliable identification of the interactions and parameters needed to realize the phase, and of the measurements needed to establish its existence~\cite{ippoliti2021manybody, GoogleTCExp}. In this context, we revisit the experimentally motivated proposal of Yao et. al.

\begin{figure*}[t!]
	\centering
	\includegraphics[width=\textwidth]{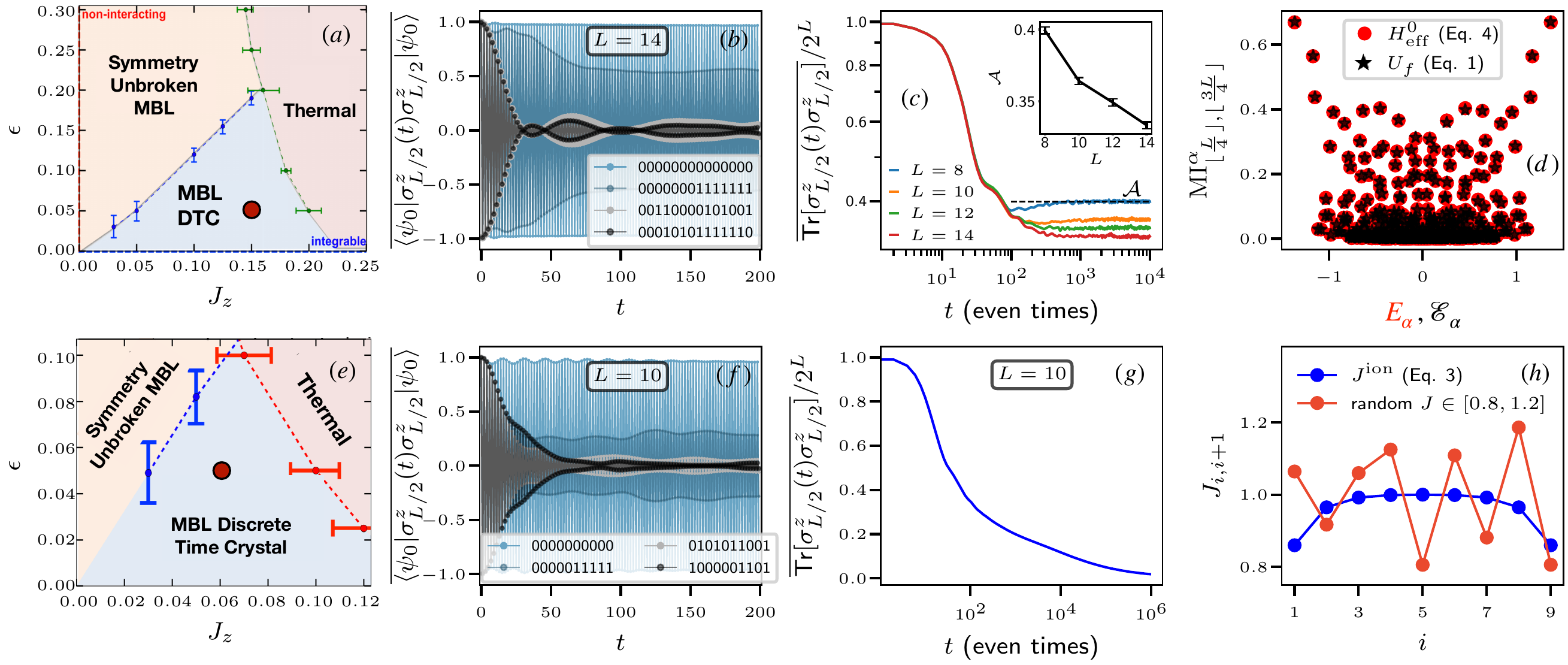}
	\caption{{\bf Absence of a discrete time-crystal in the short-range, Eq.~\eqref{eq:norm}, and long-range, Eq.~\eqref{eq:norm2}, models from Ref.~\cite{YaoDTC}}. The top panels $(a-d)$ present data for the short-range, and the bottom panels $(e-h)$ for the long-range model. $(a, e)$: Proposed phase diagrams from Ref.~\cite{YaoDTC}. The red dots indicate parameter choices deep in the proposed DTC phases, simulated in the subsequent panels. $(b, f)$: Autocorrelators for a bulk spin for different initial states in the $z$ basis, disorder averaged over 100 realizations. The data shows 
	strong initial state dependence in both models, inconsistent with the theory of MBL DTCs: oscillations from a polarized initial state last a long time with amplitude near unity, while those from a half-polarised state with one central domain wall have much lower amplitude, and those from other random states decay rapidly. $(c,g)$: Infinite temperature autocorrelators for even times, disorder averaged over $10^4$ realizations. A plateau in this quantity corresponds to a finite late-time amplitude of oscillations, $\mathcal{A}$. The short-range model displays a plateau, but this is a finite size effect with $\mathcal{A}$ decreasing with increasing $L$ (inset of $(c)$), while the long-range model only shows a decaying amplitude with no plateau, $(g)$. $(e)$: The spectrum of $U_f$ (black stars) is nearly perfectly reproduced by the leading order effective Hamiltonian, $H_{\rm eff}^0$ (red circles), which is a \emph{non-interacting, static} transverse field Ising model, Eq.~\eqref{eq:heff}, shown here for a single instance for $L=10$ (see text). $(h)$: Nearest neighbor interactions $J_{i, i+1}$ for the trapped ion experiment, Eq.~\eqref{eq:ioncouplings}, and a random instance of 20\% disorder, as in Eq.~\eqref{eq:norm}. It is not natural to model the smooth and systematic inhomogeneity of the former with the randomness of the latter, as was done in the Erratum~\cite{YaoErratum}.}
	\label{fig1}
\end{figure*}

\vspace{6pt}
\ssection{0. Models:} 
Refs.~\cite{YaoDTC, YaoErratum} study two Floquet unitaries. The first, henceforth the ``short-ranged model" reads~\cite{YaoErratum}
\begin{equation}
U_{f}= e^{-i\sum_{i}J_i^z\sigma_{i}^{z}\sigma_{i+1}^{z}+B^z_{i}\sigma_{i}^{z}}e^{-i\left(\frac{\pi}{2}-\epsilon\right)\sum_{i}\sigma_{i}^{x}},
\label{eq:norm}
\end{equation}
where $\sigma_i^{\alpha = x/y/z}$ are Pauli operators, and the couplings are drawn randomly from $B_i^z \in [0, \pi]$, $J_i^z \in [J_z - \delta J_z, J_z+\delta J_z]$, with $\delta J_z =0.2 J_z$. Ref.~\cite{YaoDTC} obtained a phase diagram for this model, reproduced in Fig.~\ref{fig1}$(a)$, by performing numerical simulations on systems of size $L \leq 14$~\footnote{The authors also show end-to-end mutual information for $L=16,18$, but this is affected by edge physics, see below. All other simulations of time-dynamics, level statistics etc. are presented for $L \leq 14$.}. 

The second, henceforth the ``long-ranged model"~\cite{YaoDTC}, is
\begin{equation}
U_{f}^{\rm ion}= e^{-i\sum_{i\neq j}J_{ij}^{\rm ion}\sigma_{i}^{z}\sigma_{j}^{z}+B^z_{i}\sigma_{i}^{z}}e^{-i\left(\frac{\pi}{2}-\epsilon\right)\sum_{i}\sigma_{i}^{x}},
\label{eq:norm2}
\end{equation}
where $B_i^z \in [0, 2\pi]$ and $J_{ij}^{\rm ion}$ models the long-range interactions in a trapped ion experimental setup~\cite{MonroeExp}. These approximately look like a power law, $J_{ij}^{\rm ion}\sim J_z/r_{ij}^\alpha$, with $\alpha \approx 1.5$. However, the combination of  trapping potential and long-range interactions results in inhomogeneities in the ion spacings, with the bulk differing from the edges. This leads to systematic deviations of the couplings from a simple power law. The exact matrix of inhomogeneous experimental couplings is given by 
\begin{equation}
    J_{ij}^{\rm ion} = \Omega^2\omega_R\sum_{m=1}^L\frac{b_{i,m} b_{j,m}}{\mu^2-\omega_m^2},
    \label{eq:ioncouplings}
\end{equation}
where $b's$ are vibrational normal modes of the ions, and the other quantities characterize the trapping potentials and normal mode frequencies. The Supplement in \cite{MonroeExp} lists the parameters used in the experimental system of $L=10$ ions. A phase diagram for this model proposed in Ref.~\cite{YaoDTC} is reproduced in Fig.~\ref{fig1}$(e)$~\footnote{The matrix of experimental couplings from Eq.~\eqref{eq:ioncouplings} was simulated to obtain the phase diagram. See footnote 53 in \cite{YaoDTC}.}. 

\vspace{6pt}
\ssection{1. Comment on choice of models:}
The long-range model is essentially {\it ab initio}.

For the short-range model, we note that its original version in Ref.~\cite{YaoDTC} did not have disorder in the $J^z$ couplings but only in the onsite $B_i$ fields (similar to the trapped ion experiment). However, this turns out to \emph{not} be sufficient to realize an MBL DTC~\cite{YaoErratum, Khemani2019, ippoliti2021manybody}~\footnote{Ising-odd terms, {\it i.e.} $B_i$, are approximately dynamically decoupled over two periods by the almost $\pi$ rotations implemented by $x$ pulses, thereby lowering their effective disorder strength and hindering localization in the absence of independent disorder in the $J^z$ couplings.}. Disorder in $J^z$ was introduced in the revised model in Eq.~\eqref{eq:norm} in an Erratum, which states ``the $\delta J_z$ disorder was originally added to emulate the spatial inhomogeneity of the coupling matrix for trapped ion experiments.''~\cite{YaoErratum}. However, the inhomogeneity of couplings in the trapped ion experiment is a \emph{systematic, non-random} edge-induced inhomogeneity. Emulating this by adding white-noise disorder in the bulk is highly unnatural. In Fig.~\ref{fig1}$(h)$ we plot the nearest neighbor ion couplings, $J_{i,i+1}^{\rm ion}$ from Eq.~(\ref{eq:ioncouplings}), together with a random instance of couplings with 20\% disorder as in Eq.~(\ref{eq:norm}), showing a stark contrast. 

However, because the short range model is also of independent interest, we set aside these issues of modeling choices, and examine the models as defined. Unfortunately we find that, for both models, the parameter regions identified as an MBL DTC in Figs.~\ref{fig1}$(a,e)$ instead belong to a thermal phase.

\vspace{6pt}
\ssection{2. Absence of stable period doubling} The temporal response of an MBL DTC is characterized by infinitely long-lived period-doubled oscillations in autocorrelators of local operators, $A_i(t) = \langle \psi_0 |\sigma_i^z(t) \sigma_i^z(0)|\psi_0\rangle$, with $t\in \mathbb{Z}$ measured in units of the driving period.  Importantly, an MBL DTC displays oscillations from  \emph{all} initial states $|\psi_0\rangle$, with no systematic state-to-state difference ~\cite{Khemani2016, Else2016, Khemani2019, ippoliti2021manybody} ~\footnote{Conceivably, there could be systems where a finite fraction of initial states lead to infinitely long lived oscillations. However, no such examples are currently known.}.  This robustness with respect to the choice of initial states is one of the constitutive features of an MBL DTC, which diagnoses the defining ``eigenstate order" in this phase, whereby \emph{every} many-body eigenstate of $U_f$ displays long range quantum correlations~\cite{Khemani2016, Else2016, vonKeyserlingk2016}; this is in contrast to a plethora of long-studied, but unrelated, dynamical phenomena which may also display period doubling, but in limited settings from special initial states~\cite{faradaywaves,Goldstein2018,Khemani2019}.
This fundamental feature of stable period doubling in an MBL DTC can be experimentally verified by systematically probing dynamics from a wide range of initial states~\cite{ippoliti2021manybody, GoogleTCExp}; for example, several quantum simulator platforms are able to initialize the system in any desired product state.

In Figs~\ref{fig1}$(b,f)$ we plot the disorder averaged autocorrelator of a bulk spin $\overline{A_{L/2}(t)}$ for different initial product states in the z-basis for the short and long ranged models in Eqs.~(\ref{eq:norm}, \ref{eq:norm2}). We simulate points in parameter space deep in the proposed MBL DTC phases: $\epsilon$ = 0.05, and $J_z=0.15/0.06 $ for the short/long ranged models respectively~\footnote{For the long-ranged model, the interaction strength $J_z$ is set by multiplying the coupling matrix in Eq.~\eqref{eq:ioncouplings} by a constant which sets the maximum nearest-neighbor interaction.}. The data shows strong initial state dependence, in contrast to the behavior of an MBL DTC: oscillations from a polarized initial state last a long time with amplitude near unity~\footnote{At these sizes, oscillations from polarized initial states last a time that scales exponentially with system size (data not shown).}, while those from a half-polarised state with one central domain wall have much lower amplitude, while oscillations from other product states with random spin configurations decay rapidly. This decay is incompatible with an MBL DTC.

 We note that Ref.~\cite{YaoDTC} only simulated a polarized initial state for the long-ranged model, which realizes atypically long oscillations. Likewise, initial state dependence was not studied in the experiment in Ref.~\cite{MonroeExp}, which only presented data for a polarized and half-polarized initial state, but at two different parameter values. However, trapped ion platforms can easily prepare different initial product states; had Ref.~\cite{MonroeExp} done so, a strong initial state dependence in thermalization rates would have been observed, implying the absence of an MBL DTC, even within the finite experimental coherence times available, $t \sim 100$, as in Fig.~\ref{fig1}$(f)$. Indeed, the trapped-ion experiment was performed on a system of only 10 ions, allowing for a confirmation of the absence of an MBL DTC in this model by numerical exact diagonalization using the experimental parameters for $J^{\rm ion}_{ij}$ and setting $2J_z=0.075$ and $\epsilon= 0.03\pi/2$ as in Ref.~\cite{MonroeExp}. This simulation was performed in Chapter 7 of \cite{Khemani2019}, showing decay of oscillations from generic initial states similar to Fig.~\ref{fig1}$(f)$.  
 We have also verified that this behavior holds at other points in the phase diagrams of both the short and long range models~\footnote{However, probing too small an $\epsilon$ is not reliable at these system sizes because the system is integrable at $\epsilon=0$.}. 
 
 Next, we present data for infinite temperature autocorrelators, $A_i^\infty(t) = \frac{1}{2^L} \mbox{Tr}[\sigma_i^z(t)\sigma_i(0)]$, whose Fourier transform was studied in Ref.~\cite{YaoDTC} for the short range model. Since $A^\infty(t)$ averages over initial states, it is less sensitive to outlier states such as the polarized one. In Fig.~\ref{fig1}$(c,g)$, we plot $\overline{A_{L/2}^\infty(t)}$ averaged over $10^4$ disorder realizations for different system sizes. We only show data for even times for clarity; a non-zero late-time plateau in this data corresponds to persistent oscillations with a finite amplitude, $\mathcal{A}$, equal to the plateau value. We find that the short-range model does display a plateau as appropriate to an an MBL DTC (Fig.~\ref{fig1}$(c)$). However, this is a finite size effect, with the amplitude $\mathcal{A}$ showing a clear decrease with increasing system size (Fig.~\ref{fig1}$(c)$, inset). This data is also strongly affected by boundary conditions, discussed next, and we find that the same parameters simulated with periodic boundary conditions, in fact, show only a steady decay with time with no visible plateau in $\overline{A_{L/2}^\infty(2t)}$ . Likewise, the long-range model does not display a plateau, consistent with a decaying amplitude of oscillations and the absence of an MBL DTC, apparent already for $L=10$ (Fig.~\ref{fig1}g).

\vspace{6pt}
\ssection{3. Boundary conditions in the short-range model:} 
The phase boundary between the MBL and thermal phases in Fig.~\ref{fig1}$(a)$ was derived in Ref.~\cite{YaoDTC} from level statistics data.
While the various temporal DTC diagnostics in \cite{YaoDTC} were obtained using {\it open} boundary conditions, we were only able to reproduce the level statistics data for {\it periodic} boundary conditions. Thus, it appears that the phase diagram in Fig.~\ref{fig1}$(a)$ was obtained by combining data with varying boundary conditions. Boundary conditions turn out to have a big impact at the small sizes $L\leq 14$ probed in Ref.~\cite{YaoDTC}, and the OBC data turns out to be strongly affected by edge physics. Unlike the bulk autocorrelators plotted in Fig.~\ref{fig1}$(b)$, autocorrelators of edge spins, $A_{1/L}(t)$ show exponentially long-lived in $L$ oscillations (at small sizes) for all initial states (data not shown). Additionally, a primary diagnostic used by Ref.~\cite{YaoDTC} was $\rm{MI}_{1, L}$, the mutual information between edge spins $(1,L)$ in eigenstates of $U_f$. In an MBL DTC, all eigenstates of $U_f$ resemble (spatially) long-range ordered ``Schr\"{o}dinger cat" states~\cite{Khemani2016, Else2016}, and $\rm{MI}_{1, L} \approx \ln(2)$ is meant to diagnose this property. However, as shown in the next section, almost all eigenstates of $U_f$, in fact, lack bulk long-range order, and the observed behavior of $\rm{MI}_{1, L}$ and $A_1(t)$
merely reflects the edge physics of an ordered transverse field Ising model. 

\vspace{6pt}
\ssection {4. What happens in the infinite size limit?} We have already shown that the system sizes and parameters probed by Ref.~\cite{YaoDTC} do not display the fundamental temporal characteristic of an MBL DTC i.e. they do not display robust period doubled oscillations.  In principle, our comment could end here. Nevertheless, as a matter of curiosity, readers may wonder whether the trends we have seen can turn around at even larger sizes, beyond those available to simulations or experiment. While the approach to the infinite size limit can be quite complicated, we provide further analysis on this question, and the observed finite size behaviors, through a high-frequency Floquet Magnus expansion. This expansion is a good approximation due to the small values of $J, \epsilon$ in Figs.~\ref{fig1}$(a,d)$. This furnishes a static effective Hamiltonian, $H_{\rm eff}$, that approximates the dynamics in pre-asymptotic regimes of finite system sizes and/or times, capturing a transient prethermal-like regime~\cite{Abanin2017, Mori2016, Else2017}. $H_{\rm eff}$ is defined over two Floquet periods as $U_f^2 \equiv e^{-2iH_{\rm eff}}$~\cite{Else2017}. 

\vspace{2pt}
\sssection{4a.\  Short-range model}
At leading order in $J,\epsilon$,
\begin{equation}
H_{\rm eff}^0 = \sum_i J_i \sigma_i^z \sigma_{i+1}^z + \epsilon \sum_i |\cos(B_i)|\sigma_i^x,
\label{eq:heff}
\end{equation}
a \emph{non-interacting} transverse field Ising model (TFIM) with Ising symmetry $P_x=\prod_i \sigma_i^x$~\footnote{See Supplementary Material in \cite{GoogleTCExp} for a derivation which reproduces the model in Eq.~\eqref{eq:heff} upto single qubit $Z$ rotations that leave the eigenvalues invariant.}. Interestingly, as shown in Fig.~\ref{fig1}$(d)$,  the spectrum of $U_f$ (black stars) is near-perfectly reproduced by $H_{\rm eff}^0$ (red circles) at the sizes under consideration. The figure displays the mutual information, ${\rm MI}^\alpha_{\floor{\frac{L}{4}}, \floor{\frac{3L}{4}}}$, between two bulk spins $(i,j)= (\floor{\frac{L}{4}}, \floor{\frac{3L}{4}})$ separated by distance $L/2$ for each eigenstate $|\alpha\rangle$ of $H_{\rm eff}^0$, plotted against its eigenvalue $E_\alpha$ (for a particular disorder instance in a system of size $L=10$). The plot also shows mutual information in the eigenstates of $U_f$ plotted against $\mathcal{E}_\alpha = i\log(\phi_\alpha^2)/2$, where $\{\phi_\alpha\}$ are eigenvalues of $U_f$; $\mathcal{E}_\alpha$ is the appropriate ``quasi-energy'' to compare with $E_\alpha$ according to  $U_f^2 \equiv e^{-2iH_{\rm eff}}$. Notably, ${\rm MI}^\alpha_{\floor{\frac{L}{4}}, \floor{\frac{3L}{4}}} \approx 0$ in most eigenstates of $U_f$, contrary to the expected spatial long-range order in an MBL DTC. A systematic analysis across disorder realizations and system sizes confirms the lack of bulk long-range order in the majority of eigenstates of $U_f$ (data not shown).

From Eq.~\eqref{eq:heff} and Fig.~\ref{fig1}$(d)$, we see that an effective non-interacting TFIM is an excellent approximation of the dynamics at small sizes. Further analysis of the non-interacting TFIM of Eq.~\eqref{eq:heff} shows that it is weakly localized with a large localization length, $\xi \sim 7$ comparable to the system sizes $L \leq 14$ amenable to exact diagonalization. At larger sizes, the system may initially look better localized as the system size grows in comparison to $\xi$. However, at larger sizes, the effects of interactions --- essential for claiming a \emph{many-body} localized-DTC ---  also assert themselves from higher order terms in the Magnus expansion. As is well known from studies of MBL, if the single particle localization length is too large, the effect of interactions is to destabilize localization~\cite{Basko2006}. More recently, theories of non-perturbative ``avalanche instabilities" of MBL predict a critical localization length, $\xi_c = 2/\ln(2) \approx 2.88$, such that an Anderson localized model with average $[\xi] > \xi_c$  (such as the system under consideration) is asymptotically unstable to the addition of interactions~\cite{deroeckAvalanche, Crowley_2020}. Thus, {not only does the short-range model lack the stable temporal dynamics of a DTC and appear non-interacting at small sizes, it is also predicted to be unstable to thermalization as interactions assert themselves at larger sizes.}

We now discuss how an effective description of the system in terms of a TFIM accounts for many finite-size observations in Ref.~\cite{YaoDTC}. The TFIM has a ground state phase transition at $J_z = \epsilon$ between a paramagnet and an ordered phase which spontaneously breaks Ising symmetry. The properties identified with a DTC in Ref.~\cite{YaoDTC}, in fact, can be accounted for by the low-energy physics of the ordered phase of $H_{\rm eff}^0$. From Eq.~\eqref{eq:norm} and Fig.~\ref{fig1}$(a)$, each $J_i^z>0$, and hence the ground state doublet of $-H_{\rm eff}^0$ in the ordered phase approximately resembles fully polarised Schr\"odinger cat states, $|\pm \rangle \approx |000\cdots 000\rangle \pm |111\cdots 111\rangle$, with exponentially small in $L$ splitting between the even and odd cats. 
\begin{itemize}[leftmargin=*]
  \setlength\itemsep{1pt}
  
    \item The DTC-paramagnet phase boundary in Fig.~\ref{fig1}(a), $J_z\sim \epsilon$, roughly tracks the ground-state phase boundary of the TFIM, with the proposed DTC  corresponding to the ordered phase.

    \item Due to long-range order in the ground states, the mutual information MI is large for low-temperature eigenstates near the edges of the spectrum, but not for excited states in the middle (Fig.~\ref{fig1}(d)). Hence most eigenstates of $U_f$ do not show bulk long-range order. 
    
    \item The presence of Majorana edge modes in the ordered phase leads to strong boundary effects. In particular, ${\rm MI}^\alpha_{1,L} \approx \ln(2)$ even for highly-excited eigenstates of a disorder-free TFIM with no bulk long-range order, directly impacting the MI diagnostic in Ref.~\cite{YaoDTC}. This also predicts long-lived autocorrelators from edge spins~\cite{Else_2017_edge}.

    \item The Poisson level statistics obtained in Ref.~\cite{YaoDTC} can reflect the integrable (non-interacting) nature of the TFIM rather than MBL. 
    
    \item Long-lived oscillations from the polarized state can be explained from the very low temperature of this state, analogous to the theory of prethermal TCs~\cite{Else2017}\footnote{We have also observed long-lived oscillations from low temperature Ne\'el states, with similar lifetimes as polarized states (data not shown).}. Strictly speaking, a prethermal TC requires ${H}_\text{eff}$ to have a symmetry breaking ordering phase transition at a finite temperature $T_c$. The magnetization, $\sigma_i^z(t)$, for initial states at temperatures $T<T_c$ relaxes to the (non-zero) equilibrium value of the order parameter at temperature $T$, which sets the amplitude of oscillations~\cite{Else2017}. However, short-ranged models in one dimension cannot have order at any finite temperature. Thus, the polarized state, which is a finite (but low) temperature state, should ultimately equilibrate to zero magnetization. However, 
    thermal correlation lengths may still exceed the system size~\footnote{In addition, for small sizes, the thermalization time to the true equilibrium value (of zero) may exceed the time set by exponential splitting of low-energy states in even/odd parity sectors, which would account for the exponential lifetime of oscillations from these low-energy states.}. This allows low temperature states to show long-lived oscillations with a finite amplitude in finite size systems, even if the equilibrium order parameter is asymptotically zero for such states. 
    Thus, while long-lived oscillations from the polarized state is reminiscent of the predictions of prethermal time-crystals~\cite{Else2017}, the model in Eq.~\eqref{eq:norm} cannot continue to display this behavior at even larger sizes due to the short range interactions. In any case, while there is plenty of interest in prethermal phenomena, this was not the focus of the claims of Ref.~\cite{YaoDTC}.

\end{itemize}

\sssection{4b.\ Long-range model} In Fig.~\ref{fig1}$(e)$, the $J$'s and $\epsilon$'s are also small enough to permit a Floquet Magnus approximation for the system sizes under consideration, with many of the same conclusions discussed above. However, due to the long-range interactions, even the leading order $H_{\rm eff}^0$ is now a strongly interacting model with no special edge physics. Importantly, long-range interactions are known to have a destabilizing effect on MBL as compared to the short-ranged case and thus there is even less reason to expect this model to be asymptotically localized if its nearest-neighbor counterpart is not. Indeed, the decay of the oscillation amplitude with time for $A^\infty(t)$ is clearly visible even for $L=10$ in Fig.~\ref{fig1}$(g)$. 
However, long-range interactions $\sim 1/r^{1.5}$ in 1D can yield a finite critical temperature $T_c$ below which the spectrum of $H_{\rm eff}$ is ordered, with the polarized state being one such ordered low temperature state. Thus, while the long-range model is not MBL, it may display the physics of a prethermal time-crystal \` a la Ref.~\cite{Else2017}, even in the limit of infinitely large sizes (if the form of the interactions remain long-range enough as the system scales up). Importantly, as mentioned earlier,  the distinction between prethermal phenomena and MBL DTCs is not just a matter of asymptotic theory, but is also experimentally observable -- even within finite experimental coherence times, $t \sim 100$ on current quantum simulators -- by systematically probing initial state dependence~\cite{ippoliti2021manybody, GoogleTCExp}.

\vspace{6pt}
\ssection {5. Summary:} Both the short-range and long-range models, Eqs.~(\ref{eq:norm},\ref{eq:norm2}), show rapid decay of oscillations from generic initial states and do not realize an MBL DTC; consequently, neither does the trapped ion experiment of Ref.~\cite{MonroeExp}.  Instead, the parameter ranges labeled as an MBL DTC in Fig.~\ref{fig1} belong to the thermal phase. Long-lived oscillations from certain initial states, such as polarized states, follow from the low-temperature physics of a static transverse field Ising model rather the nonequilibrium physics of an eigenstate ordered MBL DTC. 

We note here that the short-range model belongs to a larger family of kicked-Ising models that had been previously proposed to realize an MBL DTC~\cite{Khemani2016, Else2016, vonKeyserlingk2016}, where the phase had been found to require larger values of $J_z\sim \pi/4$ and strong $J_z$ disorder; in this regime, even the leading order term in the Magnus expansion looks interacting, and there is no small parameter controlling the expansion so all higher order terms have to be considered, making it a genuinely many-body Floquet problem already at modest system sizes. 

More generally, the observation of period doubling, by itself, has a long and celebrated history dating as far back as the work of Faraday~\cite{faradaywaves, Goldstein2018, Khemani2019}. Observing subharmonic oscillations from one or a handful of initial states a priori does not fall outside this category, and therefore does not experimentally establish an eigenstate-ordered MBL DTC phase. Rather, such a demonstration  requires a distinctive combination of observations, summarised in Refs.~\cite{Khemani2019, ippoliti2021manybody} and experimentally implemented in Ref.~\cite{GoogleTCExp}, including a systematic study of robustness to the choice of parameters and initial states, and the effects of finite system sizes and experimental coherence times.

\vspace{6pt}
\noindent \textit{Acknowledgements---} We thank Matteo Ippoliti, Curt von Keyserlingk, and Achilleas Lazarides for previous collaboration on related work. 

\bibliography{YaoComment}

%merlin.mbs apsrev4-1.bst 2010-07-25 4.21a (PWD, AO, DPC) hacked
%Control: key (0)
%Control: author (0) dotless jnrlst
%Control: editor formatted (1) identically to author
%Control: production of article title (0) allowed
%Control: page (1) range
%Control: year (0) verbatim
%Control: production of eprint (0) enabled
\begin{thebibliography}{34}%
\makeatletter
\providecommand \@ifxundefined [1]{%
 \@ifx{#1\undefined}
}%
\providecommand \@ifnum [1]{%
 \ifnum #1\expandafter \@firstoftwo
 \else \expandafter \@secondoftwo
 \fi
}%
\providecommand \@ifx [1]{%
 \ifx #1\expandafter \@firstoftwo
 \else \expandafter \@secondoftwo
 \fi
}%
\providecommand \natexlab [1]{#1}%
\providecommand \enquote  [1]{``#1''}%
\providecommand \bibnamefont  [1]{#1}%
\providecommand \bibfnamefont [1]{#1}%
\providecommand \citenamefont [1]{#1}%
\providecommand \href@noop [0]{\@secondoftwo}%
\providecommand \href [0]{\begingroup \@sanitize@url \@href}%
\providecommand \@href[1]{\@@startlink{#1}\@@href}%
\providecommand \@@href[1]{\endgroup#1\@@endlink}%
\providecommand \@sanitize@url [0]{\catcode `\\12\catcode `\$12\catcode
  `\&12\catcode `\#12\catcode `\^12\catcode `\_12\catcode `\%12\relax}%
\providecommand \@@startlink[1]{}%
\providecommand \@@endlink[0]{}%
\providecommand \url  [0]{\begingroup\@sanitize@url \@url }%
\providecommand \@url [1]{\endgroup\@href {#1}{\urlprefix }}%
\providecommand \urlprefix  [0]{URL }%
\providecommand \Eprint [0]{\href }%
\providecommand \doibase [0]{http://dx.doi.org/}%
\providecommand \selectlanguage [0]{\@gobble}%
\providecommand \bibinfo  [0]{\@secondoftwo}%
\providecommand \bibfield  [0]{\@secondoftwo}%
\providecommand \translation [1]{[#1]}%
\providecommand \BibitemOpen [0]{}%
\providecommand \bibitemStop [0]{}%
\providecommand \bibitemNoStop [0]{.\EOS\space}%
\providecommand \EOS [0]{\spacefactor3000\relax}%
\providecommand \BibitemShut  [1]{\csname bibitem#1\endcsname}%
\let\auto@bib@innerbib\@empty
%</preamble>
\bibitem [{\citenamefont {Yao}\ \emph {et~al.}(2017{\natexlab{a}})\citenamefont
  {Yao}, \citenamefont {Potter}, \citenamefont {Potirniche},\ and\
  \citenamefont {Vishwanath}}]{YaoDTC}%
  \BibitemOpen
  \bibfield  {author} {\bibinfo {author} {\bibfnamefont {N.~Y.}\ \bibnamefont
  {Yao}}, \bibinfo {author} {\bibfnamefont {A.~C.}\ \bibnamefont {Potter}},
  \bibinfo {author} {\bibfnamefont {I.-D.}\ \bibnamefont {Potirniche}}, \ and\
  \bibinfo {author} {\bibfnamefont {A.}~\bibnamefont {Vishwanath}},\ }\bibfield
   {title} {\enquote {\bibinfo {title} {Discrete time crystals: Rigidity,
  criticality, and realizations},}\ }\href {\doibase
  10.1103/PhysRevLett.118.030401} {\bibfield  {journal} {\bibinfo  {journal}
  {Phys. Rev. Lett.}\ }\textbf {\bibinfo {volume} {118}},\ \bibinfo {pages}
  {030401} (\bibinfo {year} {2017}{\natexlab{a}})}\BibitemShut {NoStop}%
\bibitem [{\citenamefont {Yao}\ \emph {et~al.}(2017{\natexlab{b}})\citenamefont
  {Yao}, \citenamefont {Potter}, \citenamefont {Potirniche},\ and\
  \citenamefont {Vishwanath}}]{YaoErratum}%
  \BibitemOpen
  \bibfield  {author} {\bibinfo {author} {\bibfnamefont {N.~Y.}\ \bibnamefont
  {Yao}}, \bibinfo {author} {\bibfnamefont {A.~C.}\ \bibnamefont {Potter}},
  \bibinfo {author} {\bibfnamefont {I.-D.}\ \bibnamefont {Potirniche}}, \ and\
  \bibinfo {author} {\bibfnamefont {A.}~\bibnamefont {Vishwanath}},\ }\bibfield
   {title} {\enquote {\bibinfo {title} {Erratum: Discrete time crystals:
  Rigidity, criticality, and realizations [phys. rev. lett. 118, 030401
  (2017)]},}\ }\href {\doibase 10.1103/PhysRevLett.118.269901} {\bibfield
  {journal} {\bibinfo  {journal} {Phys. Rev. Lett.}\ }\textbf {\bibinfo
  {volume} {118}},\ \bibinfo {pages} {269901} (\bibinfo {year}
  {2017}{\natexlab{b}})}\BibitemShut {NoStop}%
\bibitem [{\citenamefont {Zhang}\ \emph {et~al.}(2017)\citenamefont {Zhang},
  \citenamefont {Hess}, \citenamefont {Kyprianidis}, \citenamefont {Becker},
  \citenamefont {Lee}, \citenamefont {Smith}, \citenamefont {Pagano},
  \citenamefont {Potirniche}, \citenamefont {Potter}, \citenamefont
  {Vishwanath}, \citenamefont {Yao},\ and\ \citenamefont {Monroe}}]{MonroeExp}%
  \BibitemOpen
  \bibfield  {author} {\bibinfo {author} {\bibfnamefont {J.}~\bibnamefont
  {Zhang}}, \bibinfo {author} {\bibfnamefont {P.~W.}\ \bibnamefont {Hess}},
  \bibinfo {author} {\bibfnamefont {A.}~\bibnamefont {Kyprianidis}}, \bibinfo
  {author} {\bibfnamefont {P.}~\bibnamefont {Becker}}, \bibinfo {author}
  {\bibfnamefont {A.}~\bibnamefont {Lee}}, \bibinfo {author} {\bibfnamefont
  {J.}~\bibnamefont {Smith}}, \bibinfo {author} {\bibfnamefont
  {G.}~\bibnamefont {Pagano}}, \bibinfo {author} {\bibfnamefont {I.-D.}\
  \bibnamefont {Potirniche}}, \bibinfo {author} {\bibfnamefont {A.~C.}\
  \bibnamefont {Potter}}, \bibinfo {author} {\bibfnamefont {A.}~\bibnamefont
  {Vishwanath}}, \bibinfo {author} {\bibfnamefont {N.~Y.}\ \bibnamefont {Yao}},
  \ and\ \bibinfo {author} {\bibfnamefont {C.}~\bibnamefont {Monroe}},\
  }\bibfield  {title} {\enquote {\bibinfo {title} {Observation of a discrete
  time crystal},}\ }\href {\doibase 10.1038/nature21413} {\bibfield  {journal}
  {\bibinfo  {journal} {Nature}\ }\textbf {\bibinfo {volume} {543}},\ \bibinfo
  {pages} {217--220} (\bibinfo {year} {2017})}\BibitemShut {NoStop}%
\bibitem [{\citenamefont {Khemani}\ \emph {et~al.}(2016)\citenamefont
  {Khemani}, \citenamefont {Lazarides}, \citenamefont {Moessner},\ and\
  \citenamefont {Sondhi}}]{Khemani2016}%
  \BibitemOpen
  \bibfield  {author} {\bibinfo {author} {\bibfnamefont {V.}~\bibnamefont
  {Khemani}}, \bibinfo {author} {\bibfnamefont {A.}~\bibnamefont {Lazarides}},
  \bibinfo {author} {\bibfnamefont {R.}~\bibnamefont {Moessner}}, \ and\
  \bibinfo {author} {\bibfnamefont {S.}~\bibnamefont {Sondhi}},\ }\bibfield
  {title} {\enquote {\bibinfo {title} {Phase structure of driven quantum
  systems},}\ }\href {\doibase 10.1103/PhysRevLett.116.250401} {\bibfield
  {journal} {\bibinfo  {journal} {Phys. Rev. Lett.}\ }\textbf {\bibinfo
  {volume} {116}},\ \bibinfo {pages} {250401} (\bibinfo {year}
  {2016})}\BibitemShut {NoStop}%
\bibitem [{\citenamefont {Else}\ \emph {et~al.}(2016)\citenamefont {Else},
  \citenamefont {Bauer},\ and\ \citenamefont {Nayak}}]{Else2016}%
  \BibitemOpen
  \bibfield  {author} {\bibinfo {author} {\bibfnamefont {Dominic~V.}\
  \bibnamefont {Else}}, \bibinfo {author} {\bibfnamefont {Bela}\ \bibnamefont
  {Bauer}}, \ and\ \bibinfo {author} {\bibfnamefont {Chetan}\ \bibnamefont
  {Nayak}},\ }\bibfield  {title} {\enquote {\bibinfo {title} {Floquet time
  crystals},}\ }\href {\doibase 10.1103/PhysRevLett.117.090402} {\bibfield
  {journal} {\bibinfo  {journal} {Phys. Rev. Lett.}\ }\textbf {\bibinfo
  {volume} {117}},\ \bibinfo {pages} {090402} (\bibinfo {year}
  {2016})}\BibitemShut {NoStop}%
\bibitem [{\citenamefont {von Keyserlingk}\ \emph {et~al.}(2016)\citenamefont
  {von Keyserlingk}, \citenamefont {Khemani},\ and\ \citenamefont
  {Sondhi}}]{vonKeyserlingk2016}%
  \BibitemOpen
  \bibfield  {author} {\bibinfo {author} {\bibfnamefont {C.~W.}\ \bibnamefont
  {von Keyserlingk}}, \bibinfo {author} {\bibfnamefont {Vedika}\ \bibnamefont
  {Khemani}}, \ and\ \bibinfo {author} {\bibfnamefont {S.~L.}\ \bibnamefont
  {Sondhi}},\ }\bibfield  {title} {\enquote {\bibinfo {title} {Absolute
  stability and spatiotemporal long-range order in floquet systems},}\ }\href
  {\doibase 10.1103/PhysRevB.94.085112} {\bibfield  {journal} {\bibinfo
  {journal} {Phys. Rev. B}\ }\textbf {\bibinfo {volume} {94}},\ \bibinfo
  {pages} {085112} (\bibinfo {year} {2016})}\BibitemShut {NoStop}%
\bibitem [{\citenamefont {Mi}\ \emph {et~al.}(2021)\citenamefont {Mi},
  \citenamefont {Ippoliti}, \citenamefont {Quintana}, \citenamefont {Greene},
  \citenamefont {Chen}, \citenamefont {Gross} \emph {et~al.}}]{GoogleTCExp}%
  \BibitemOpen
  \bibfield  {author} {\bibinfo {author} {\bibfnamefont {Xiao}\ \bibnamefont
  {Mi}}, \bibinfo {author} {\bibfnamefont {Matteo}\ \bibnamefont {Ippoliti}},
  \bibinfo {author} {\bibfnamefont {Chris}\ \bibnamefont {Quintana}}, \bibinfo
  {author} {\bibfnamefont {Amy}\ \bibnamefont {Greene}}, \bibinfo {author}
  {\bibfnamefont {Zijun}\ \bibnamefont {Chen}}, \bibinfo {author}
  {\bibfnamefont {Jonathan}\ \bibnamefont {Gross}},  \emph {et~al.},\
  }\href@noop {} {\enquote {\bibinfo {title} {Observation of time-crystalline
  eigenstate order on a quantum processor},}\ } (\bibinfo {year} {2021}),\
  \Eprint {http://arxiv.org/abs/2107.13571} {arXiv:2107.13571 [quant-ph]}
  \BibitemShut {NoStop}%
\bibitem [{\citenamefont {Randall}\ \emph {et~al.}(2021)\citenamefont
  {Randall}, \citenamefont {Bradley}, \citenamefont {van~der Gronden},
  \citenamefont {Galicia}, \citenamefont {Abobeih}, \citenamefont {Markham},
  \citenamefont {Twitchen}, \citenamefont {Machado}, \citenamefont {Yao},\ and\
  \citenamefont {Taminiau}}]{DelftTCExp}%
  \BibitemOpen
  \bibfield  {author} {\bibinfo {author} {\bibfnamefont {J.}~\bibnamefont
  {Randall}}, \bibinfo {author} {\bibfnamefont {C.~E.}\ \bibnamefont
  {Bradley}}, \bibinfo {author} {\bibfnamefont {F.~V.}\ \bibnamefont {van~der
  Gronden}}, \bibinfo {author} {\bibfnamefont {A.}~\bibnamefont {Galicia}},
  \bibinfo {author} {\bibfnamefont {M.~H.}\ \bibnamefont {Abobeih}}, \bibinfo
  {author} {\bibfnamefont {M.}~\bibnamefont {Markham}}, \bibinfo {author}
  {\bibfnamefont {D.~J.}\ \bibnamefont {Twitchen}}, \bibinfo {author}
  {\bibfnamefont {F.}~\bibnamefont {Machado}}, \bibinfo {author} {\bibfnamefont
  {N.~Y.}\ \bibnamefont {Yao}}, \ and\ \bibinfo {author} {\bibfnamefont
  {T.~H.}\ \bibnamefont {Taminiau}},\ }\href@noop {} {\enquote {\bibinfo
  {title} {Observation of a many-body-localized discrete time crystal with a
  programmable spin-based quantum simulator},}\ } (\bibinfo {year} {2021}),\
  \Eprint {http://arxiv.org/abs/2107.00736} {arXiv:2107.00736 [quant-ph]}
  \BibitemShut {NoStop}%
\bibitem [{\citenamefont {Frey}\ and\ \citenamefont
  {Rachel}(2021)}]{RachelTCExp}%
  \BibitemOpen
  \bibfield  {author} {\bibinfo {author} {\bibfnamefont {Philipp}\ \bibnamefont
  {Frey}}\ and\ \bibinfo {author} {\bibfnamefont {Stephan}\ \bibnamefont
  {Rachel}},\ }\href@noop {} {\enquote {\bibinfo {title} {Simulating a discrete
  time crystal over 57 qubits on a quantum computer},}\ } (\bibinfo {year}
  {2021}),\ \Eprint {http://arxiv.org/abs/2105.06632} {arXiv:2105.06632
  [quant-ph]} \BibitemShut {NoStop}%
\bibitem [{\citenamefont {Xu}\ \emph {et~al.}(2021)\citenamefont {Xu},
  \citenamefont {Zhang}, \citenamefont {Han}, \citenamefont {Li}, \citenamefont
  {Xue}, \citenamefont {Liu}, \citenamefont {Jin},\ and\ \citenamefont
  {Yu}}]{BeijingTCExp}%
  \BibitemOpen
  \bibfield  {author} {\bibinfo {author} {\bibfnamefont {Huikai}\ \bibnamefont
  {Xu}}, \bibinfo {author} {\bibfnamefont {Jingning}\ \bibnamefont {Zhang}},
  \bibinfo {author} {\bibfnamefont {Jiaxiu}\ \bibnamefont {Han}}, \bibinfo
  {author} {\bibfnamefont {Zhiyuan}\ \bibnamefont {Li}}, \bibinfo {author}
  {\bibfnamefont {Guangming}\ \bibnamefont {Xue}}, \bibinfo {author}
  {\bibfnamefont {Weiyang}\ \bibnamefont {Liu}}, \bibinfo {author}
  {\bibfnamefont {Yirong}\ \bibnamefont {Jin}}, \ and\ \bibinfo {author}
  {\bibfnamefont {Haifeng}\ \bibnamefont {Yu}},\ }\href@noop {} {\enquote
  {\bibinfo {title} {Realizing discrete time crystal in an one-dimensional
  superconducting qubit chain},}\ } (\bibinfo {year} {2021}),\ \Eprint
  {http://arxiv.org/abs/2108.00942} {arXiv:2108.00942 [quant-ph]} \BibitemShut
  {NoStop}%
\bibitem [{\citenamefont {Choi}\ \emph {et~al.}(2017)\citenamefont {Choi},
  \citenamefont {Choi}, \citenamefont {Landig}, \citenamefont {Kucsko},
  \citenamefont {Zhou}, \citenamefont {Isoya}, \citenamefont {Jelezko},
  \citenamefont {Onoda}, \citenamefont {Sumiya}, \citenamefont {Khemani},
  \citenamefont {von Keyserlingk}, \citenamefont {Yao}, \citenamefont
  {Demler},\ and\ \citenamefont {Lukin}}]{LukinExp}%
  \BibitemOpen
  \bibfield  {author} {\bibinfo {author} {\bibfnamefont {Soonwon}\ \bibnamefont
  {Choi}}, \bibinfo {author} {\bibfnamefont {Joonhee}\ \bibnamefont {Choi}},
  \bibinfo {author} {\bibfnamefont {Renate}\ \bibnamefont {Landig}}, \bibinfo
  {author} {\bibfnamefont {Georg}\ \bibnamefont {Kucsko}}, \bibinfo {author}
  {\bibfnamefont {Hengyun}\ \bibnamefont {Zhou}}, \bibinfo {author}
  {\bibfnamefont {Junichi}\ \bibnamefont {Isoya}}, \bibinfo {author}
  {\bibfnamefont {Fedor}\ \bibnamefont {Jelezko}}, \bibinfo {author}
  {\bibfnamefont {Shinobu}\ \bibnamefont {Onoda}}, \bibinfo {author}
  {\bibfnamefont {Hitoshi}\ \bibnamefont {Sumiya}}, \bibinfo {author}
  {\bibfnamefont {Vedika}\ \bibnamefont {Khemani}}, \bibinfo {author}
  {\bibfnamefont {Curt}\ \bibnamefont {von Keyserlingk}}, \bibinfo {author}
  {\bibfnamefont {Norman~Y.}\ \bibnamefont {Yao}}, \bibinfo {author}
  {\bibfnamefont {Eugene}\ \bibnamefont {Demler}}, \ and\ \bibinfo {author}
  {\bibfnamefont {Mikhail~D.}\ \bibnamefont {Lukin}},\ }\bibfield  {title}
  {\enquote {\bibinfo {title} {Observation of discrete time-crystalline order
  in a disordered dipolar many-body system},}\ }\href {\doibase
  10.1038/nature21426} {\bibfield  {journal} {\bibinfo  {journal} {Nature}\
  }\textbf {\bibinfo {volume} {543}},\ \bibinfo {pages} {221--225} (\bibinfo
  {year} {2017})}\BibitemShut {NoStop}%
\bibitem [{\citenamefont {Pal}\ \emph {et~al.}(2018)\citenamefont {Pal},
  \citenamefont {Nishad}, \citenamefont {Mahesh},\ and\ \citenamefont
  {Sreejith}}]{Sreejith}%
  \BibitemOpen
  \bibfield  {author} {\bibinfo {author} {\bibfnamefont {Soham}\ \bibnamefont
  {Pal}}, \bibinfo {author} {\bibfnamefont {Naveen}\ \bibnamefont {Nishad}},
  \bibinfo {author} {\bibfnamefont {T.~S.}\ \bibnamefont {Mahesh}}, \ and\
  \bibinfo {author} {\bibfnamefont {G.~J.}\ \bibnamefont {Sreejith}},\
  }\bibfield  {title} {\enquote {\bibinfo {title} {Temporal order in
  periodically driven spins in star-shaped clusters},}\ }\href {\doibase
  10.1103/PhysRevLett.120.180602} {\bibfield  {journal} {\bibinfo  {journal}
  {Phys. Rev. Lett.}\ }\textbf {\bibinfo {volume} {120}},\ \bibinfo {pages}
  {180602} (\bibinfo {year} {2018})}\BibitemShut {NoStop}%
\bibitem [{\citenamefont {Rovny}\ \emph {et~al.}(2018)\citenamefont {Rovny},
  \citenamefont {Blum},\ and\ \citenamefont {Barrett}}]{BarrettExp}%
  \BibitemOpen
  \bibfield  {author} {\bibinfo {author} {\bibfnamefont {Jared}\ \bibnamefont
  {Rovny}}, \bibinfo {author} {\bibfnamefont {Robert~L.}\ \bibnamefont {Blum}},
  \ and\ \bibinfo {author} {\bibfnamefont {Sean~E.}\ \bibnamefont {Barrett}},\
  }\bibfield  {title} {\enquote {\bibinfo {title} {Observation of
  discrete-time-crystal signatures in an ordered dipolar many-body system},}\
  }\href {\doibase 10.1103/PhysRevLett.120.180603} {\bibfield  {journal}
  {\bibinfo  {journal} {Phys. Rev. Lett.}\ }\textbf {\bibinfo {volume} {120}},\
  \bibinfo {pages} {180603} (\bibinfo {year} {2018})}\BibitemShut {NoStop}%
\bibitem [{\citenamefont {Khemani}\ \emph {et~al.}(2019)\citenamefont
  {Khemani}, \citenamefont {Moessner},\ and\ \citenamefont
  {Sondhi}}]{Khemani2019}%
  \BibitemOpen
  \bibfield  {author} {\bibinfo {author} {\bibfnamefont {V.}~\bibnamefont
  {Khemani}}, \bibinfo {author} {\bibfnamefont {R.}~\bibnamefont {Moessner}}, \
  and\ \bibinfo {author} {\bibfnamefont {S.}~\bibnamefont {Sondhi}},\
  }\bibfield  {title} {\enquote {\bibinfo {title} {A brief history of time
  crystals},}\ }\href {https://arxiv.org/abs/1910.10745} {\bibfield  {journal}
  {\bibinfo  {journal} {arXiv e-prints}\ ,\ \bibinfo {pages}
  {arXiv:1910.10745}} (\bibinfo {year} {2019})}\BibitemShut {NoStop}%
\bibitem [{\citenamefont {Ippoliti}\ \emph {et~al.}(2021)\citenamefont
  {Ippoliti}, \citenamefont {Kechedzhi}, \citenamefont {Moessner},
  \citenamefont {Sondhi},\ and\ \citenamefont
  {Khemani}}]{ippoliti2021manybody}%
  \BibitemOpen
  \bibfield  {author} {\bibinfo {author} {\bibfnamefont {Matteo}\ \bibnamefont
  {Ippoliti}}, \bibinfo {author} {\bibfnamefont {Kostyantyn}\ \bibnamefont
  {Kechedzhi}}, \bibinfo {author} {\bibfnamefont {Roderich}\ \bibnamefont
  {Moessner}}, \bibinfo {author} {\bibfnamefont {S.~L.}\ \bibnamefont
  {Sondhi}}, \ and\ \bibinfo {author} {\bibfnamefont {Vedika}\ \bibnamefont
  {Khemani}},\ }\href@noop {} {\enquote {\bibinfo {title} {Many-body physics in
  the nisq era: quantum programming a discrete time crystal},}\ } (\bibinfo
  {year} {2021}),\ \Eprint {http://arxiv.org/abs/2007.11602} {arXiv:2007.11602
  [cond-mat.dis-nn]} \BibitemShut {NoStop}%
\bibitem [{Note1()}]{Note1}%
  \BibitemOpen
  \bibinfo {note} {The authors also show end-to-end mutual information for
  $L=16,18$, but this is affected by edge physics, see below. All other
  simulations of time-dynamics, level statistics etc. are presented for $L \leq
  14$.}\BibitemShut {Stop}%
\bibitem [{Note2()}]{Note2}%
  \BibitemOpen
  \bibinfo {note} {The matrix of experimental couplings from Eq.~\protect
  \textup {\hbox {\mathsurround \z@ \protect \normalfont (\ignorespaces \ref
  {eq:ioncouplings}\unskip \@@italiccorr )}} was simulated to obtain the phase
  diagram. See footnote 53 in \cite {YaoDTC}.}\BibitemShut {Stop}%
\bibitem [{Note3()}]{Note3}%
  \BibitemOpen
  \bibinfo {note} {Ising-odd terms, {\protect \it i.e.} $B_i$, are
  approximately dynamically decoupled over two periods by the almost $\pi $
  rotations implemented by $x$ pulses, thereby lowering their effective
  disorder strength and hindering localization in the absence of independent
  disorder in the $J^z$ couplings.}\BibitemShut {Stop}%
\bibitem [{Note4()}]{Note4}%
  \BibitemOpen
  \bibinfo {note} {Conceivably, there could be systems where a finite fraction
  of initial states lead to infinitely long lived oscillations. However, no
  such examples are currently known.}\BibitemShut {Stop}%
\bibitem [{\citenamefont {Faraday}(1831)}]{faradaywaves}%
  \BibitemOpen
  \bibfield  {author} {\bibinfo {author} {\bibfnamefont {M.}~\bibnamefont
  {Faraday}},\ }\bibfield  {title} {\enquote {\bibinfo {title} {On a peculiar
  class of acoustical figures; and on certain forms assumed by groups of
  particles upon vibrating elastic surfaces},}\ }\href
  {http://www.jstor.org/stable/107936} {\bibfield  {journal} {\bibinfo
  {journal} {Philosophical Transactions of the Royal Society of London}\
  }\textbf {\bibinfo {volume} {121}},\ \bibinfo {pages} {299--340} (\bibinfo
  {year} {1831})}\BibitemShut {NoStop}%
\bibitem [{\citenamefont {{Goldstein}}(2018)}]{Goldstein2018}%
  \BibitemOpen
  \bibfield  {author} {\bibinfo {author} {\bibfnamefont {Raymond~E.}\
  \bibnamefont {{Goldstein}}},\ }\bibfield  {title} {\enquote {\bibinfo {title}
  {{Coffee stains, cell receptors, and time crystals: Lessons from the old
  literature}},}\ }\href {\doibase 10.1063/PT.3.4019} {\bibfield  {journal}
  {\bibinfo  {journal} {Physics Today}\ }\textbf {\bibinfo {volume} {71}},\
  \bibinfo {pages} {32--38} (\bibinfo {year} {2018})},\ \Eprint
  {http://arxiv.org/abs/1811.08179} {arXiv:1811.08179 [cond-mat.soft]}
  \BibitemShut {NoStop}%
\bibitem [{Note5()}]{Note5}%
  \BibitemOpen
  \bibinfo {note} {For the long-ranged model, the interaction strength $J_z$ is
  set by multiplying the coupling matrix in Eq.~\protect \textup {\hbox
  {\mathsurround \z@ \protect \normalfont (\ignorespaces \ref
  {eq:ioncouplings}\unskip \@@italiccorr )}} by a constant which sets the
  maximum nearest-neighbor interaction.}\BibitemShut {Stop}%
\bibitem [{Note6()}]{Note6}%
  \BibitemOpen
  \bibinfo {note} {At these sizes, oscillations from polarized initial states
  last a time that scales exponentially with system size (data not
  shown).}\BibitemShut {Stop}%
\bibitem [{Note7()}]{Note7}%
  \BibitemOpen
  \bibinfo {note} {However, probing too small an $\epsilon $ is not reliable at
  these system sizes because the system is integrable at $\epsilon
  =0$.}\BibitemShut {Stop}%
\bibitem [{\citenamefont {Abanin}\ \emph {et~al.}(2017)\citenamefont {Abanin},
  \citenamefont {De~Roeck}, \citenamefont {Ho},\ and\ \citenamefont
  {Huveneers}}]{Abanin2017}%
  \BibitemOpen
  \bibfield  {author} {\bibinfo {author} {\bibfnamefont {D.}~\bibnamefont
  {Abanin}}, \bibinfo {author} {\bibfnamefont {W.}~\bibnamefont {De~Roeck}},
  \bibinfo {author} {\bibfnamefont {W.}~\bibnamefont {Ho}}, \ and\ \bibinfo
  {author} {\bibfnamefont {F.}~\bibnamefont {Huveneers}},\ }\bibfield  {title}
  {\enquote {\bibinfo {title} {A rigorous theory of many-body prethermalization
  for periodically driven and closed quantum systems},}\ }\href {\doibase
  10.1007/s00220-017-2930-x} {\bibfield  {journal} {\bibinfo  {journal}
  {Commun. Math. Phys.}\ }\textbf {\bibinfo {volume} {354}},\ \bibinfo {pages}
  {809--827} (\bibinfo {year} {2017})}\BibitemShut {NoStop}%
\bibitem [{\citenamefont {Mori}\ \emph {et~al.}(2016)\citenamefont {Mori},
  \citenamefont {Kuwahara},\ and\ \citenamefont {Saito}}]{Mori2016}%
  \BibitemOpen
  \bibfield  {author} {\bibinfo {author} {\bibfnamefont {T.}~\bibnamefont
  {Mori}}, \bibinfo {author} {\bibfnamefont {T.}~\bibnamefont {Kuwahara}}, \
  and\ \bibinfo {author} {\bibfnamefont {K.}~\bibnamefont {Saito}},\ }\bibfield
   {title} {\enquote {\bibinfo {title} {Rigorous bound on energy absorption and
  generic relaxation in periodically driven quantum systems},}\ }\href
  {\doibase 10.1103/PhysRevLett.116.120401} {\bibfield  {journal} {\bibinfo
  {journal} {Phys. Rev. Lett.}\ }\textbf {\bibinfo {volume} {116}},\ \bibinfo
  {pages} {120401} (\bibinfo {year} {2016})}\BibitemShut {NoStop}%
\bibitem [{\citenamefont {Else}\ \emph
  {et~al.}(2017{\natexlab{a}})\citenamefont {Else}, \citenamefont {Bauer},\
  and\ \citenamefont {Nayak}}]{Else2017}%
  \BibitemOpen
  \bibfield  {author} {\bibinfo {author} {\bibfnamefont {D.}~\bibnamefont
  {Else}}, \bibinfo {author} {\bibfnamefont {B.}~\bibnamefont {Bauer}}, \ and\
  \bibinfo {author} {\bibfnamefont {C.}~\bibnamefont {Nayak}},\ }\bibfield
  {title} {\enquote {\bibinfo {title} {Prethermal phases of matter protected by
  time-translation symmetry},}\ }\href {\doibase 10.1103/PhysRevX.7.011026}
  {\bibfield  {journal} {\bibinfo  {journal} {Phys. Rev. X}\ }\textbf {\bibinfo
  {volume} {7}},\ \bibinfo {pages} {011026} (\bibinfo {year}
  {2017}{\natexlab{a}})}\BibitemShut {NoStop}%
\bibitem [{Note8()}]{Note8}%
  \BibitemOpen
  \bibinfo {note} {See Supplementary Material in \cite {GoogleTCExp} for a
  derivation which reproduces the model in Eq.~\protect \textup {\hbox
  {\mathsurround \z@ \protect \normalfont (\ignorespaces \ref {eq:heff}\unskip
  \@@italiccorr )}} upto single qubit $Z$ rotations that leave the eigenvalues
  invariant.}\BibitemShut {Stop}%
\bibitem [{\citenamefont {Basko}\ \emph {et~al.}(2006)\citenamefont {Basko},
  \citenamefont {Aleiner},\ and\ \citenamefont {Altshuler}}]{Basko2006}%
  \BibitemOpen
  \bibfield  {author} {\bibinfo {author} {\bibfnamefont {D.~M.}\ \bibnamefont
  {Basko}}, \bibinfo {author} {\bibfnamefont {I.~L.}\ \bibnamefont {Aleiner}},
  \ and\ \bibinfo {author} {\bibfnamefont {B.~L.}\ \bibnamefont {Altshuler}},\
  }\bibfield  {title} {\enquote {\bibinfo {title} {Metal-insulator transition
  in a weakly interacting many-electron system with localized single-particle
  states},}\ }\href {\doibase 10.1016/j.aop.2005.11.014} {\bibfield  {journal}
  {\bibinfo  {journal} {Ann. Phys. (Amsterdam)}\ }\textbf {\bibinfo {volume}
  {321}},\ \bibinfo {pages} {1126--1205} (\bibinfo {year} {2006})}\BibitemShut
  {NoStop}%
\bibitem [{\citenamefont {De~Roeck}\ and\ \citenamefont
  {Huveneers}(2017)}]{deroeckAvalanche}%
  \BibitemOpen
  \bibfield  {author} {\bibinfo {author} {\bibfnamefont {Wojciech}\
  \bibnamefont {De~Roeck}}\ and\ \bibinfo {author} {\bibfnamefont {Fran\ifmmode
  \mbox{\c{c}}\else~\c{c}\fi{}ois}\ \bibnamefont {Huveneers}},\ }\bibfield
  {title} {\enquote {\bibinfo {title} {Stability and instability towards
  delocalization in many-body localization systems},}\ }\href {\doibase
  10.1103/PhysRevB.95.155129} {\bibfield  {journal} {\bibinfo  {journal} {Phys.
  Rev. B}\ }\textbf {\bibinfo {volume} {95}},\ \bibinfo {pages} {155129}
  (\bibinfo {year} {2017})}\BibitemShut {NoStop}%
\bibitem [{\citenamefont {Crowley}\ and\ \citenamefont
  {Chandran}(2020)}]{Crowley_2020}%
  \BibitemOpen
  \bibfield  {author} {\bibinfo {author} {\bibfnamefont {P.~J.~D.}\
  \bibnamefont {Crowley}}\ and\ \bibinfo {author} {\bibfnamefont
  {A.}~\bibnamefont {Chandran}},\ }\bibfield  {title} {\enquote {\bibinfo
  {title} {Avalanche induced coexisting localized and thermal regions in
  disordered chains},}\ }\href {\doibase 10.1103/physrevresearch.2.033262}
  {\bibfield  {journal} {\bibinfo  {journal} {Physical Review Research}\
  }\textbf {\bibinfo {volume} {2}} (\bibinfo {year} {2020}),\
  10.1103/physrevresearch.2.033262}\BibitemShut {NoStop}%
\bibitem [{\citenamefont {Else}\ \emph
  {et~al.}(2017{\natexlab{b}})\citenamefont {Else}, \citenamefont {Fendley},
  \citenamefont {Kemp},\ and\ \citenamefont {Nayak}}]{Else_2017_edge}%
  \BibitemOpen
  \bibfield  {author} {\bibinfo {author} {\bibfnamefont {Dominic~V.}\
  \bibnamefont {Else}}, \bibinfo {author} {\bibfnamefont {Paul}\ \bibnamefont
  {Fendley}}, \bibinfo {author} {\bibfnamefont {Jack}\ \bibnamefont {Kemp}}, \
  and\ \bibinfo {author} {\bibfnamefont {Chetan}\ \bibnamefont {Nayak}},\
  }\bibfield  {title} {\enquote {\bibinfo {title} {Prethermal strong zero modes
  and topological qubits},}\ }\href {\doibase 10.1103/physrevx.7.041062}
  {\bibfield  {journal} {\bibinfo  {journal} {Physical Review X}\ }\textbf
  {\bibinfo {volume} {7}} (\bibinfo {year} {2017}{\natexlab{b}}),\
  10.1103/physrevx.7.041062}\BibitemShut {NoStop}%
\bibitem [{Note9()}]{Note9}%
  \BibitemOpen
  \bibinfo {note} {We have also observed long-lived oscillations from low
  temperature Ne\'el states, with similar lifetimes as polarized states (data
  not shown).}\BibitemShut {Stop}%
\bibitem [{Note10()}]{Note10}%
  \BibitemOpen
  \bibinfo {note} {In addition, for small sizes, the thermalization time to the
  true equilibrium value (of zero) may exceed the time set by exponential
  splitting of low-energy states in even/odd parity sectors, which would
  account for the exponential lifetime of oscillations from these low-energy
  states.}\BibitemShut {Stop}%
\end{thebibliography}%
\end{document}